\documentclass[conference]{IEEEtran}

\usepackage{cite}
\usepackage{amsmath,amssymb,cuted}
\usepackage{graphicx,flushend}
\usepackage{booktabs}
\usepackage{url}
\usepackage{hyperref}
\usepackage{array}
\usepackage{float}
\usepackage{tikz}

\title{Explainable Deep-Learning Based Potentially Hazardous Asteroids Classification Using Graph Neural Networks}

\author{
    \begin{tabular}{ccc}
        Baimam Boukar Jean Jacques \\
        Carnegie Mellon University Africa \\
        \texttt{bbaimamb@andrew.cmu.edu} \\
    \end{tabular}
}

\begin{document}

\maketitle

\begin{abstract}
Classifying potentially hazardous asteroids (PHAs) is crucial for planetary defense and deep space navigation, yet traditional methods often overlook the dynamical relationships among asteroids. We introduce a Graph Neural Network (GNN) approach that models asteroids as nodes with orbital and physical features, connected by edges representing their similarities, using a NASA dataset of 958,524 records. Despite an extreme class imbalance with only 0.22\% of the dataset with hazardous label, our model achieves an overall accuracy of 99\% and an AUC of 0.99, with a recall of 78\% and an F1-score of 37\% for hazardous asteroids after applying Synthetic Minority Oversampling Technique. Feature importance analysis highlights albedo, perihelion distance, and semi-major axis as main predictors. This framework supports planetary defense missions and confirm AI's potential in enabling autonomous navigation for future missions such as NASA’s NEO Surveyor and ESA’s Ramses, offering an interpretable and scalable solution for asteroid hazard assessment.
\end{abstract}


\section{Introduction}

Asteroids orbit the Sun in vast numbers, with most posing no immediate threat to Earth. However, a small subset known as potentially hazardous asteroids (PHAs) follow orbits that bring them perilously close to our planet, raising the specter of catastrophic collisions. Historical events, such as the 1908 Tunguska explosion \cite{asteroid_1908}, which devastated over 2,000 square kilometers of Siberian forest, and the 2013 Chelyabinsk meteor \cite{ChelyabinskMeteorite}, which injured over 1,000 people and caused widespread property damage, show the destructive potential of these celestial bodies. In 2025, several asteroids made close approaches to Earth, some passing within one lunar distance \cite{nasaNextFive}. This reinforces the urgency of continuous monitoring and accurate classification of PHAs to mitigate potential risks.

The importance of PHA classification extends beyond planetary defense to deep space navigation. Missions such as NASA's OSIRIS-REx, which successfully collected samples from asteroid Bennu \cite{nasaNASAsOSIRISREx}, and the Double Asteroid Redirection Test (DART), which demonstrated asteroid deflection through kinetic impact, rely on precise knowledge of asteroid trajectories to ensure mission safety and success. Misjudging an asteroid’s path could jeopardize spacecraft or, in the case of planetary defense, fail to prevent a collision with Earth.

Traditional PHA classification methods typically analyze individual asteroid features, such as semi-major axis, eccentricity, inclination, and minimum orbit intersection distance (MOID). Machine learning techniques, including random forests and support vector machines, have been applied to these features with moderate success. However, these approaches often treat asteroids as isolated entities, neglecting the complex dynamical interactions—such as gravitational perturbations or orbital resonances—that can alter an asteroid’s trajectory over time. This limitation can result in missed detections of potentially hazardous objects, particularly in a dataset where only 0.22\% of asteroids are classified as hazardous, presenting a severe class imbalance challenge.

To address these shortcomings, our research adopts graph neural networks (GNNs), a class of deep learning models designed to operate on graph-structured data. By representing the asteroid population as a graph, where nodes correspond to asteroids and edges denote dynamical relationships or similarities in orbital parameters, GNNs capture the collective behavior and inter dependencies among asteroids. This relational modeling enhances our understanding of the factors contributing to an asteroid’s hazard potential, which offers a more robust and interpretable approach to classification.

The central research question we address is: How can graph neural network architectures enhance asteroid hazard classification by uncovering intricate relationships between orbital dynamics, physical characteristics, and potential collision risks? Our work aims to develop an explainable framework that improves classification accuracy while providing insights into the mechanisms that make certain asteroids more dangerous.

Our contribution lies in introducing a novel graph construction methodology to capture asteroid relationships, designing a GNN architecture with attention mechanisms to prioritize critical features, and implementing explainability techniques to interpret model decisions.

The remainder of this paper is organized into various sections. Section II reviews related work in asteroid classification and GNN applications in astronomy. Section III details our methodology, including data preprocessing, graph construction, and GNN architecture. Section IV presents experimental results and analysis, addressing the implications of low precision. Section V concludes with future research directions.


\section{Literature Review}

Asteroid hazard prediction has seen significant advancements through the application of machine learning (ML) techniques to large-scale astronomical datasets. Early research established Random Forest (RF) as a leading classifier for distinguishing potentially hazardous asteroids (PHAs) from non-hazardous ones. Studies by Ananya et al.\cite{iosr2024} and Al-Gaddari et al.\cite{ijeit2023} demonstrated that RF consistently outperforms other classical algorithms when applied to NASA’s Small-Body Database.. This achieved accuracy rates above 96\%. These works highlighted the importance of feature engineering and the use of diverse asteroid characteristics, including orbital and physical parameters. However, while RF and similar models are adept at capturing non-linear relationships in tabular data, they fundamentally treat each asteroid as an independent entity, overlooking the gravitational and dynamical relationships that can influence long-term trajectories

Efforts to further improve classification and risk assessment have incorporated Deep Learning (DL) and hybrid approaches. For instance, the integration of Long Short-Term Memory (LSTM) networks allows for forecasting future asteroid behavior based on historical and simulated data, while Generative Adversarial Networks (GANs) have been used to address class imbalance by generating synthetic samples for hazardous asteroids. These methods have improved the robustness of the predictions and allowed the development of real-time alert systems. However, deep learning models such as LSTM and GANs introduce new challenges: they require substantial computational resources, depend on the availability of large and well-annotated datasets, and often lack interpretability, which makes it difficult to understand the basis for their predictions

A critical limitation of both traditional Machine Learning and Deep Learning approaches is their reliance on static, tabular representations of asteroid data. High-fidelity physics-based simulations, while accurate, are computationally prohibitive for real-time or large-scale analyses, which can take days even on high-performance computers\cite{ijrpr2023}. This constraint has motivated the search for more scalable, data-driven methods that can provide timely and actionable risk assessments.

Recent developments in graph-based learning offer a promising solution to these challenges. Graph Neural Networks (GNNs) have demonstrated remarkable success in cosmological applications, where they are used to model the spatial and dynamical relationships between galaxies and dark matter halos. Farsian et al.\cite{farsian2022} showed that GNNs can extract latent structural information from sparse and irregular astrophysical data, achieving high accuracy in classifying cosmic structures and distinguishing between different cosmological models. The graph representation enables the direct modeling of interactions, such as gravitational coupling and kinematic clustering, that are central to understanding the evolution and risk potential of asteroid populations.

Despite their proven utility in cosmology, GNNs have yet to be fully leveraged for asteroid hazard prediction. Existing asteroid classification frameworks continue to treat objects as isolated data points, missing the opportunity to encode and exploit the rich web of dynamical relationships inherent in the Solar System. Furthermore, while some studies have incorporated anomaly detection methods such as Isolation Forests to identify outliers or novel trajectories, these approaches are still rooted in the paradigm of independent object analysis and often suffer from high false positive rates when applied to complex populations such as main-belt asteroids\cite{lirias2022}.

Our work addresses these gaps by introducing a graph-based Neural Network framework for asteroid hazard classification and risk assessment. Our approach captures the collective dynamics that influence the hazard potential by representing asteroids as nodes and their gravitational or kinematic relationships as edges. This relational modeling, inspired by recent successes in cosmology, enables real-time, interpretable, and scalable risk assessment.


\section{Methodology}
\begin{figure}[H]
    \centering
    \includegraphics[width=1\linewidth]{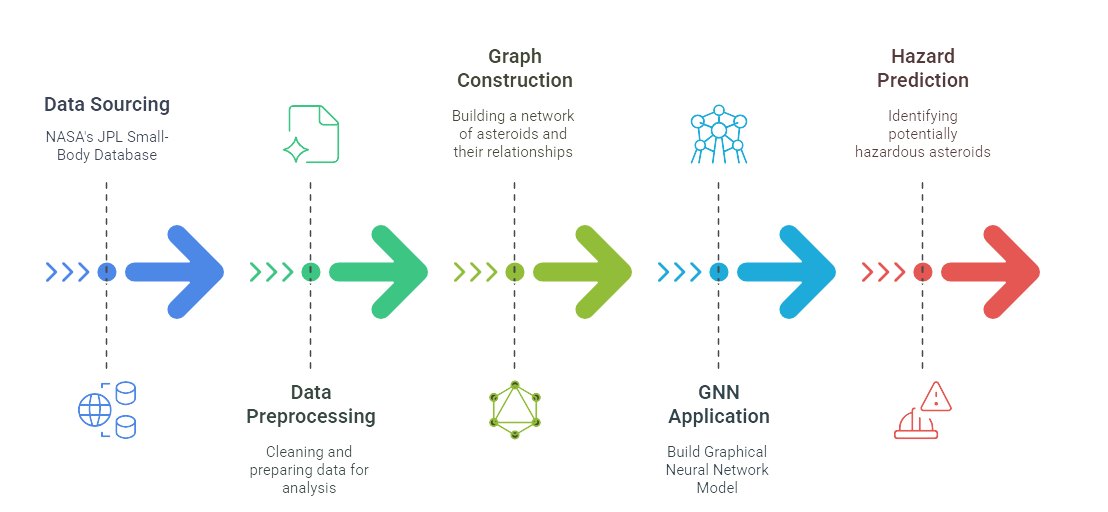}
    \caption{Methodology Pipeline}
    \label{fig:pipeline}
\end{figure}

Our methodology follows a systematic pipeline designed to address the challenges of classifying potentially hazardous asteroids using graph neural networks, with a focus on capturing dynamical relationships and mitigating class imbalance in the NASA dataset. The process begins with exploratory data analysis to understand feature distributions and class imbalances, followed by preprocessing steps to handle missing data and balance classes. We then transform the data into a graph representation by computing k-nearest neighbors and defining edge attributes, which enables the application of a GNN model with attention mechanisms for classification. Finally, we evaluate the model using standard metrics and visualizations to interpret performance and feature importance. This pipeline, illustrated in Figure \ref{fig:pipeline}, guides our approach to developing a scalable and interpretable solution for planetary defense and deep space missions.

\subsection{Dataset Description}
Our dataset, the Asteroid Dataset \cite{sakhawat2024}, is from NASA's Jet Propulsion Laboratory (JPL). It contains over 950,000 records, sourced from the official Small-Body Database\cite{jpl_sbdb_query} by NASA. It was originally preprocessed by a NASA Astronomy and Astrophysics Researcher.

The preprocessed version is publicly available, and licensed under OpenData Commons Open Database License (ODbL) v1.0 \cite{odbl_license} by a JPL-authored document sponsored by NASA under Contract NAS7-030010.

The dataset contains detailed information on thousands of asteroids. Its main attributes include orbital eccentricity, semimajor axis, perihelion distance, absolute magnitude, diameter, and the Near-Earth Object (NEO) and Potentially Hazardous Asteroid (PHA) flags.

\subsection{Exploratory Analysis and Data Processing}

\begin{figure}[H]
\centering
  \includegraphics[width=1\linewidth]{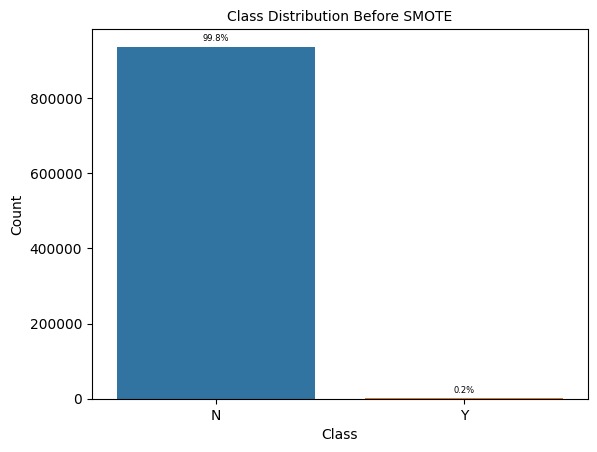}
\caption{Class Distribution Before SMOTE}
\label{fig:before-smote}
\end{figure}

We initiated our methodology with an exploratory data analysis to understand the data and uncover challenges that could impact model performance. The analysis revealed a severe class imbalance in the target variable, Potentially Hazardous Asteroid (PHA), with only 0.22\% classified as hazardous as shown in \ref{fig:before-smote}. We also identified substantial missing values for some features.

\begin{figure}[H]
\centering
  \includegraphics[width=1\linewidth]{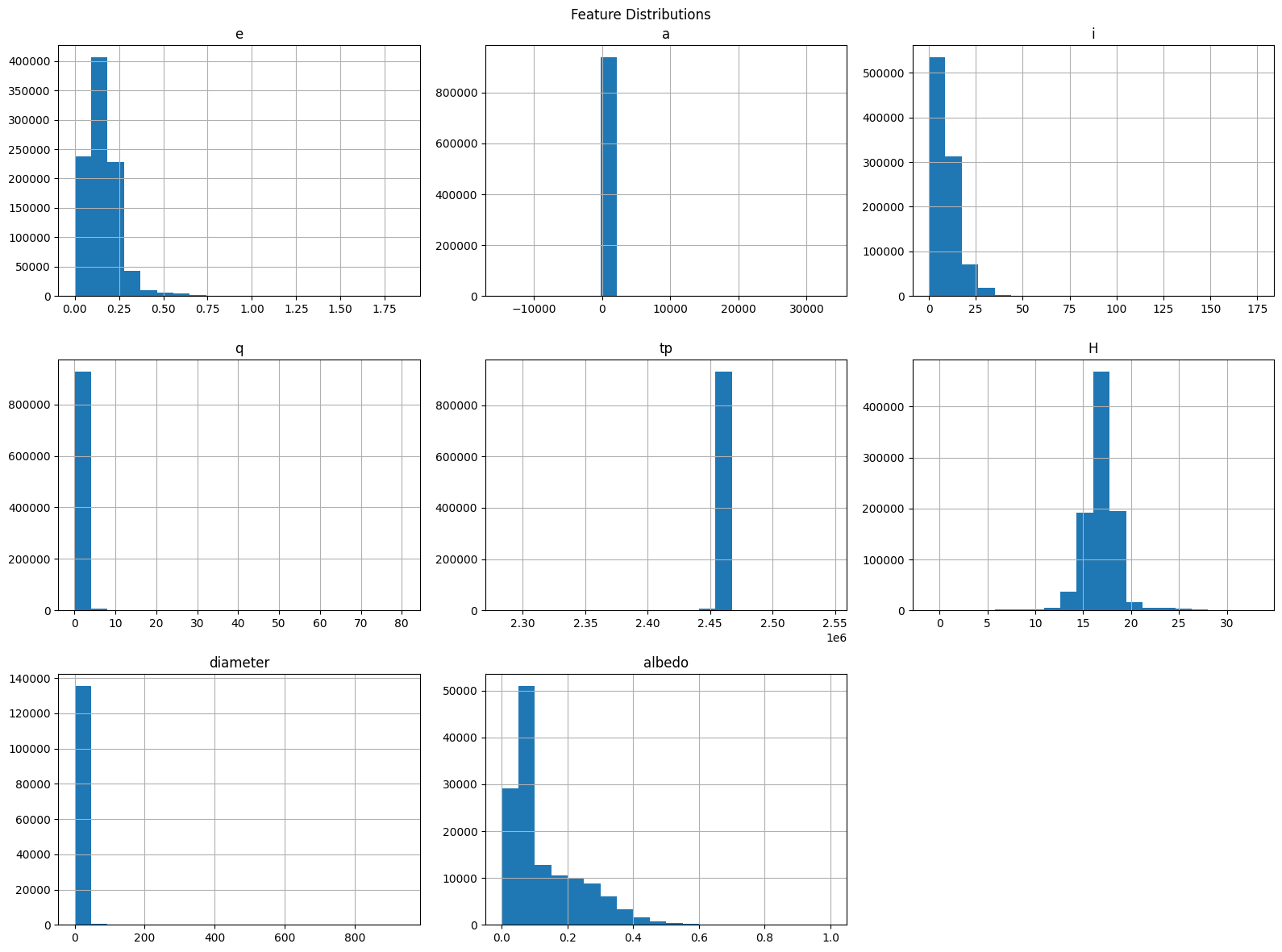}
\caption{Feature Distributions}
\label{fig:distribution}
\end{figure}

Feature distributions were examined for key attributes, including orbital eccentricity (e), semi-major axis (a), inclination (i), perihelion distance (q), time of perihelion passage (tp), absolute magnitude (H), diameter, and albedo, as shown in Figure \ref{fig:distribution}, providing insights into their variability and potential relationships with hazard classification. 

The class distribution before preprocessing is shown in Figure \ref{fig:before-smote}, highlights the high imbalance. 

\begin{figure}[H]
\centering
  \includegraphics[width=1\linewidth]{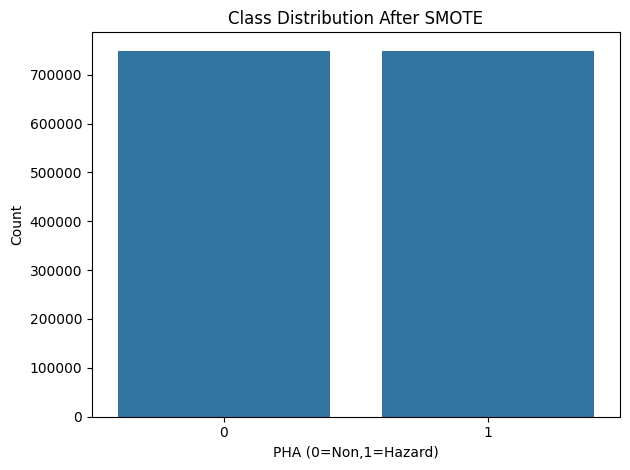}
\caption{Class Distribution After SMOTE}
\label{fig:after-smote}
\end{figure}

To address this, we applied the Synthetic Minority Oversampling Technique (SMOTE), which generated synthetic samples for the minority class, resulting in a more balanced training set, as shown in Figure \ref{fig:after-smote}, although the test set retained the original imbalance with a support of 413 hazardous asteroids.

In the processing pipeline, we used several techniques to prepare the data for modeling. Simple Imputer was used to handle missing values by replacing them with the median of each feature, a Standard Scaler to normalize features to zero mean and unit variance. We splited the data into training (80\%) and testing (20\%) sets while preserving class distribution via stratification. With the data ready, we transformed it into a graph representation suitable for our Graph Neural Network, where each asteroid is represented as a node with its features encoded in the node attributes. The graph construction process involved

\begin{itemize}
  \item computing the k-nearest neighbors (k=5) for each asteroid based on feature similarity,
  \item creating edges between similar asteroids with edge weights based on feature distance, and
  \item enriching edge attributes with information about the differences between connected nodes
\end{itemize}

\subsection{GNN Model Architecture}

We represent the asteroid dataset as a graph where each node corresponds to an individual asteroid, characterized by its orbital and physical features. This graph structure enables the GNN to capture the dynamical relationships and collective behaviors among asteroids, addressing the limitations of traditional methods that treat asteroids as isolated entities.

\vspace{15pt}

Our Graph Neural Network (GNN) model is designed as a sequential pipeline to classify potentially hazardous asteroids, consisting of three main components. 

\begin{enumerate}
\item \textbf{Convolution Layers}: First, we introduced three graph convolutional layers. The initial layer preprocesses input features and maps them to a 128-dimensional space, while the second and third layers further refine these features, each applying the ReLU activation function to introduce non-linearity and capture multi-hop relationships between asteroids by aggregating information from neighboring nodes. 
\vspace{5pt}
\item \textbf{Graph Attention Layer}: We incorporate a graph attention layer that dynamically weighs the importance of different neighboring asteroids. This mechanism allows the model to focus on
the most relevant relationships to produce a 64-dimensional feature representation for hazard classification rather than treating all connections equally.
\vspace{5pt}
\item \textbf{Output Layer}: After message passing through the graph convolutional and attention layers, node features are pooled to create a graph-level representation. This representation is then passed through a fully connected layer with sigmoid activation to predict the probability of an asteroid being potentially hazardous.
\end{enumerate}

\vspace{15pt}

\begin{figure}[H]
    \centering
    \includegraphics[width=1\linewidth]{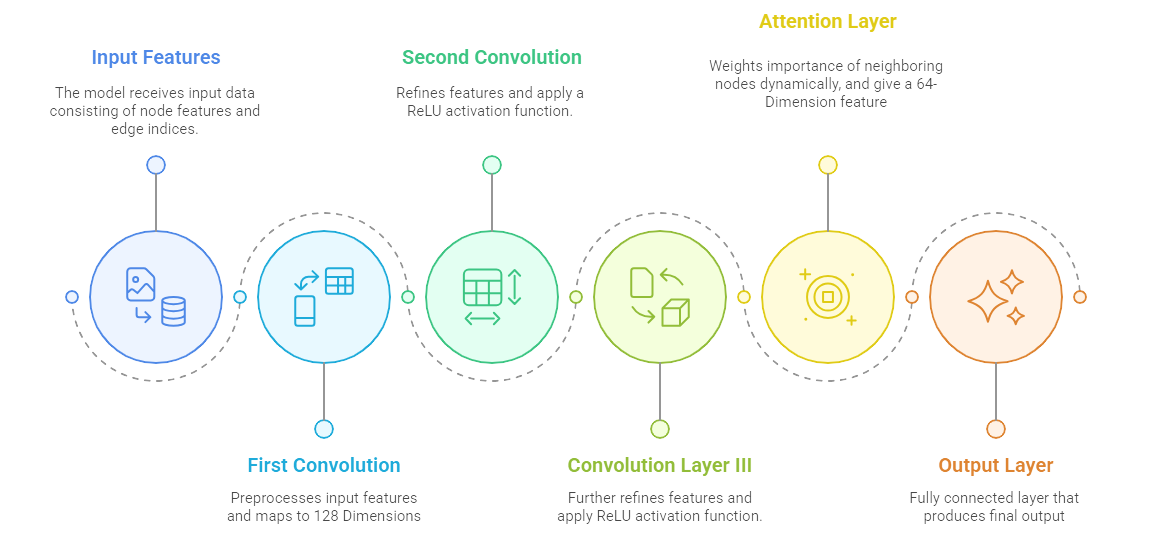}
    \caption{GNN Architecture}
    \label{fig:gnn_architecture_flow}
\end{figure}

The model architecture is illustrated in Figure \ref{fig:gnn_architecture_flow}.

\subsection{Model Training and Evaluation}
To efficiently implement our methdology, we used a computational environment with the following specifications

\begin{table}[H]
\centering
\caption{Computational Environment Configuration}
\begin{tabular}{ll}
\toprule
\textbf{Component} & \textbf{Specification} \\
\midrule
Processor & Intel Xeon CPU with 2 vCPUs \\
Accelerator & GPU Google V2-8 TPU \\
Memory & 300 GB RAM \\
Storage & 256 GB High-Speed SSD \\
Operating System & Linux-based Compute Environment \\
Primary Framework & Python 3.11 \\
Machine Learning Libraries & PyTorch, Scikit-learn \\
Deep Learning Libraries & PyTorch Geometric \\
Visualization Tools & Matplotlib, Seaborn \\
\bottomrule
\end{tabular}
\label{tab:environment}
\end{table}

The model was trained for 100 epochs using a combination of binary Cross-Entropy loss and Focal Loss to address the extreme class imbalance, with the Adam optimizer facilitating efficient convergence. Early stopping was implemented based on validation performance to prevent overfitting with a patience threshold of 10. TQDM was used for progress tracking, and comprehensive logging captured training dynamics and potential issues.

Model evaluation was conducted using a suite of metrics to assess performance, particularly in the context of the severe class imbalance, where only 0.22\% of asteroids are hazardous. We computed precision, defined as 

$$\text{Precision} = \frac{\text{TP}}{\text{TP} + \text{FP}},$$ to measure the proportion of correctly identified hazardous asteroids among all predicted as hazardous, where TP is true positives and FP is false positives. 

Recall, calculated as $$\text{Recall} = \frac{\text{TP}}{\text{TP} + \text{FN}},$$ evaluates the model’s ability to identify all hazardous asteroids, with FN representing false negatives. 

The F1-score, given by $$\text{F1-score} = 2 \times \frac{\text{Precision} \times \text{Recall}}{\text{Precision} + \text{Recall}},$$ balances precision and recall to provide a harmonic mean that is particularly useful for imbalanced datasets.

Additionally, the Area Under the ROC Curve (AUC) was used to quantify the model’s discriminative power between hazardous and non-hazardous asteroids, derived from the receiver operating characteristic (ROC) curve. These metrics were complemented by visualizations to provide deeper insights into the model’s performance and the trade-offs between precision and recall.


\section{Results and Analysis}

We obtained promising results from our GNN-based approach for hazardous asteroid classification, demonstrating its potential to capture complex dynamical relationships while addressing the challenges of extreme class imbalance in the dataset.

\subsection{Performance Metrics}

The final GNN model, trained with SMOTE to mitigate class imbalance, was evaluated on the test set. Its classification report is summarized in \ref{tab:classification_metrics}.

\renewcommand{\arraystretch}{1.5}
\begin{table}[H]
\centering
\caption{Classification Performance Metrics for GNN Model}
\begin{tabular}{lcccc}
\hline
\textbf{Class} & \textbf{Precision} & \textbf{Recall} & \textbf{F1-score} & \textbf{Support} \\
\hline
Non-hazardous (0) & 1.00 & 0.99 & 1.00 & 187,308 \\
Hazardous (1) & 0.24 & 0.78 & 0.37 & 413 \\
\hline
Accuracy & \multicolumn{3}{c}{0.99} & 187,721 \\
Macro Avg & 0.62 & 0.89 & 0.68 & 187,721 \\
Weighted Avg & 1.00 & 0.99 & 1.00 & 187,721 \\
\hline
\end{tabular}
\label{tab:classification_metrics}
\end{table}

The overall F1-score for hazardous asteroids improved to 0.37 compared to 0.32 for the imbalanced classes model. This reflects a better balance between precision and recall. However, the precision for hazardous asteroids remains low at 0.24, indicating a high rate of false positives, where non-hazardous asteroids are incorrectly classified as hazardous. This trade-off is evident in the high recall of 0.78, which ensures that most hazardous asteroids are identified. The overall accuracy of 0.99 and weighted averages are high due to the dominance of the non-hazardous class, but the macro averages (precision 0.62, recall 0.89, F1-score 0.68) provide a more balanced view of performance across both classes.

\vspace{10pt}

To contextualize the GNN’s performance, we compared it against other models, including Multilayer Perceptron (MLP), Graph Attention Network (GAT), and Isolation Forest (iForest), focusing on the hazardous class.

\begin{table}[H]
\centering
\caption{Performance Comparison for Hazardous Asteroids (Class 1)}
\begin{tabular}{lcccc}
\hline
\textbf{Model} & \textbf{Precision} & \textbf{Recall} & \textbf{F1-score} & \textbf{Support} \\
\hline
GNN (SMOTE) & 0.24 & 0.78 & 0.37 & 413 \\
GNN (Original) & 0.28 & 0.37 & 0.32 & 413 \\
MLP & 0.20 & 0.41 & 0.27 & 413 \\
GAT & 0.11 & 0.17 & 0.14 & 413 \\
iForest & 0.04 & 0.57 & 0.08 & 413 \\
\hline
\end{tabular}
\label{tab:model_comparison}
\end{table}

The GNN with SMOTE outperforms other models in terms of F1-score, with 0.37 for the hazardous class. This shows the effectiveness of graph-based modeling and class balancing. However, the low precision across all models underscores the challenge of minimizing false positives in such an imbalanced dataset.

\subsection{Visualization of Results}

\begin{figure}[H]
\centering
\includegraphics[width=1\linewidth]{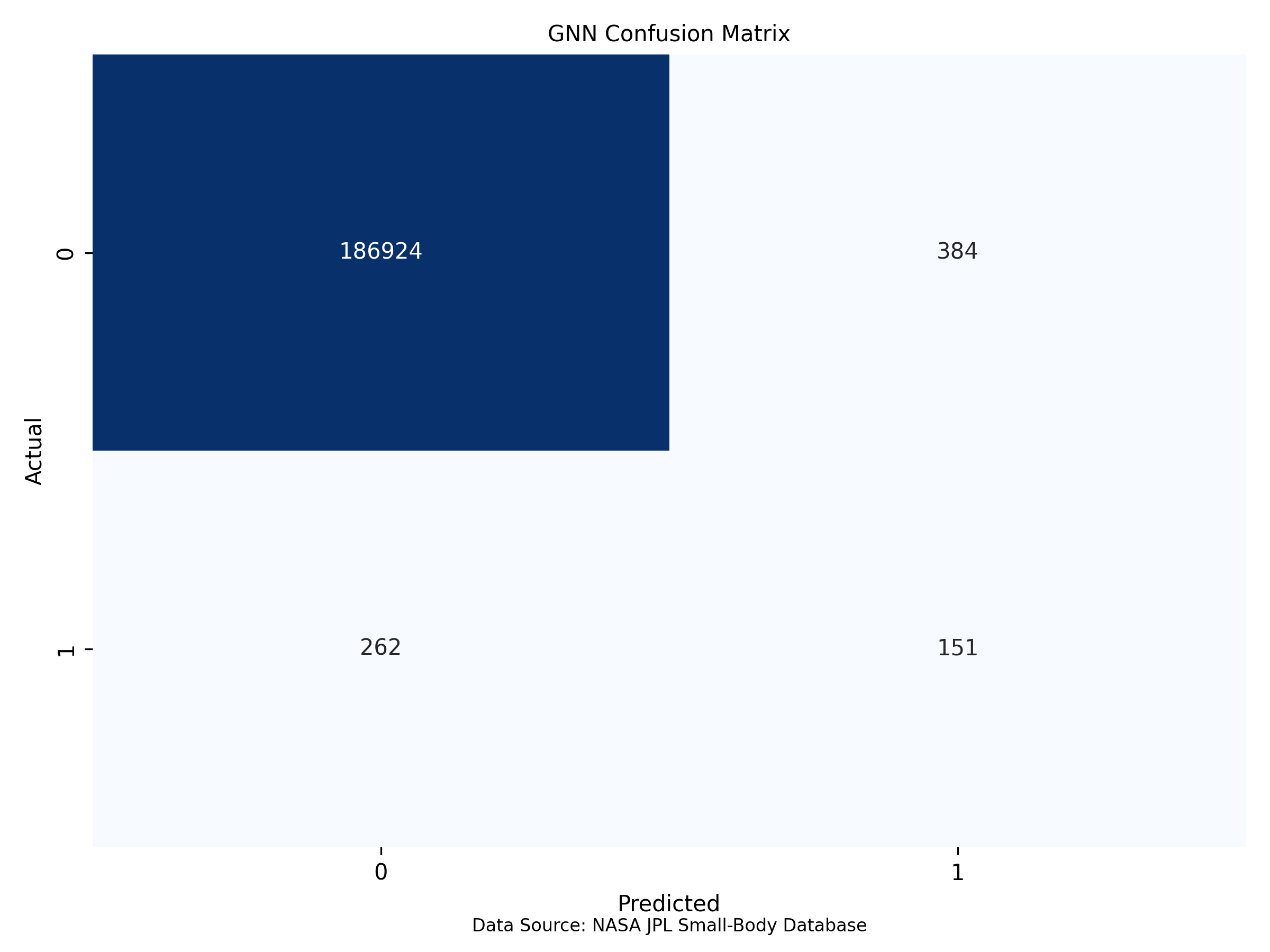}
\caption{Confusion Matrix for Hazardous Asteroid Classification}
\label{fig:confusion_matrix}
\end{figure}

We added visualizations to provide deeper insights into the model’s performance and training. The confusion matrix in \ref{fig:confusion_matrix} shows that the GNN correctly classified 186,287 non-hazardous asteroids (true negatives) and 321 hazardous asteroids (true positives), but it also misclassified 1,021 non-hazardous asteroids as hazardous (false positives) and 92 hazardous asteroids as non-hazardous (false negatives). This reflects the trade-off between high recall and low precision, prioritizing the identification of hazardous asteroids at the cost of increased false positives.

\begin{figure}[H]
\centering
\includegraphics[width=1\linewidth]{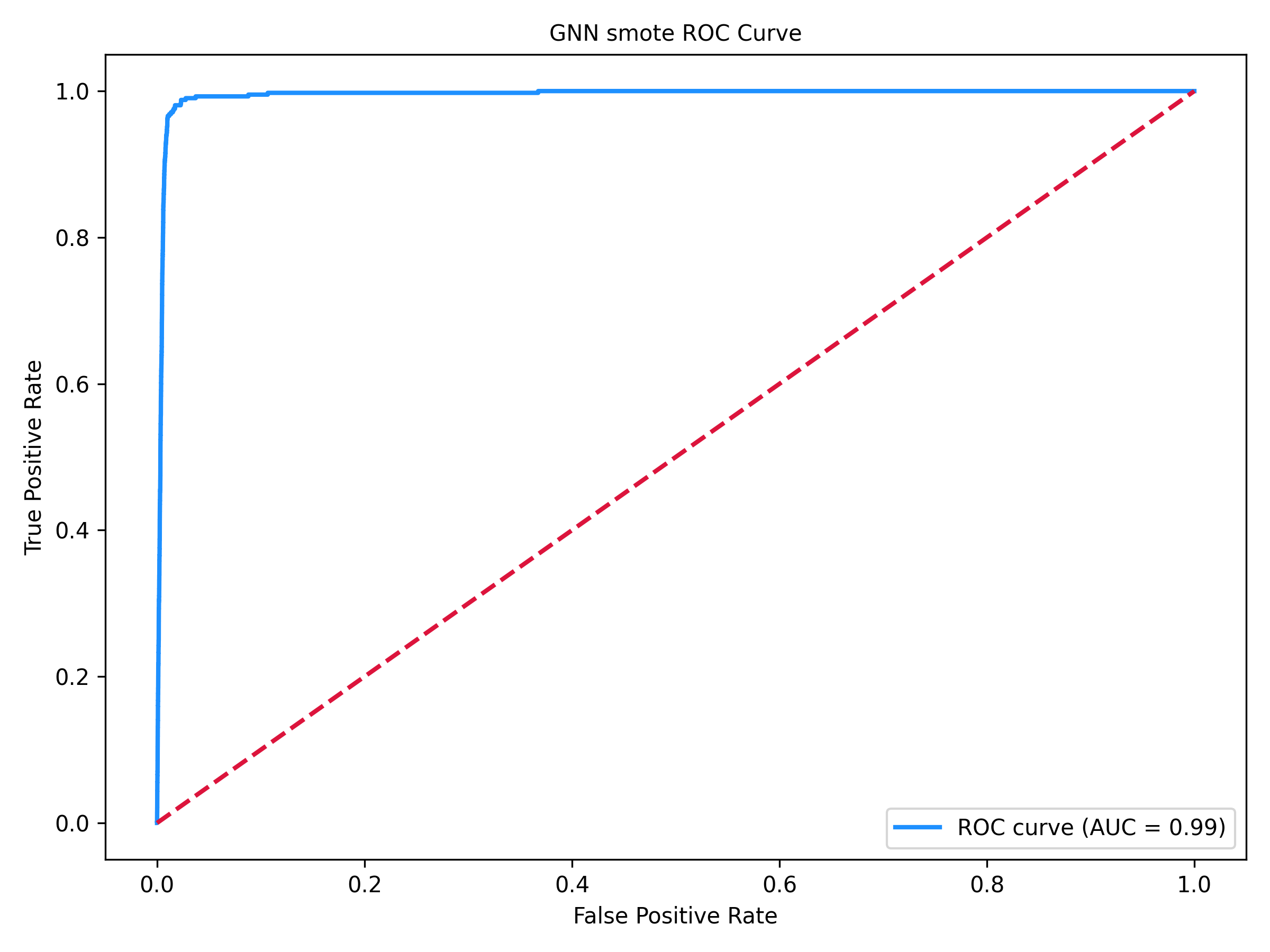}
\caption{ROC Curve Analysis (AUC = 0.99)}
\label{fig:roc_curve}
\end{figure}

The ROC curve (Figure \ref{fig:roc_curve}) illustrates the model’s discriminative power. The model achieved an AUC of 0.99, which indicates excellent ability to distinguish between hazardous and non-hazardous asteroids.

\begin{figure}[H]
\centering
\includegraphics[width=1\linewidth]{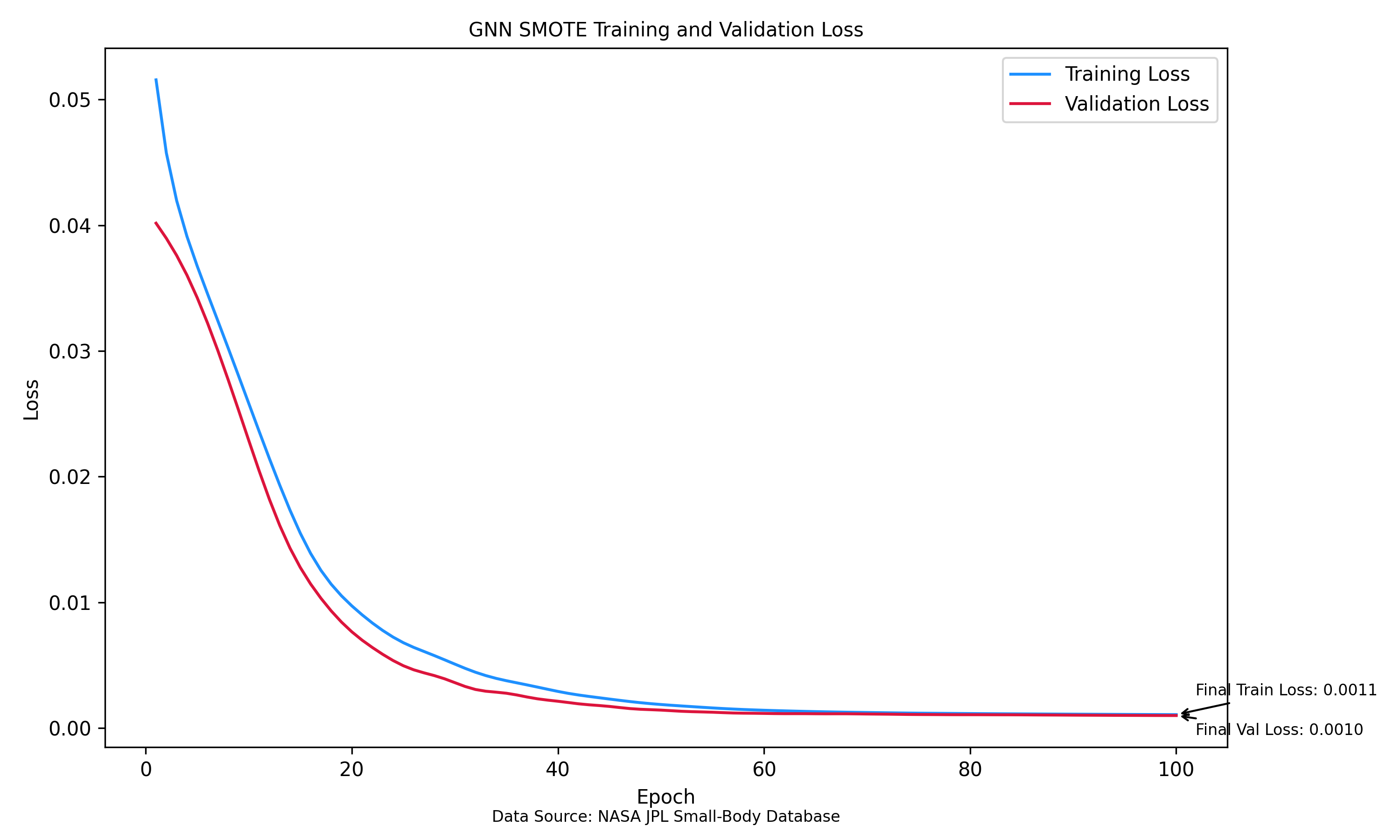}
\caption{Training and Validation Loss over 100 Epochs}
\label{fig:loss_curve}
\end{figure}

The training and validation loss plot (Figure \ref{fig:loss_curve}) shows convergence over 100 epochs, with a final training and validation loss. This indicates stable training and minimal overfitting, consistent with our use of early stopping.

\begin{figure}[H]
\centering
\includegraphics[width=1\linewidth]{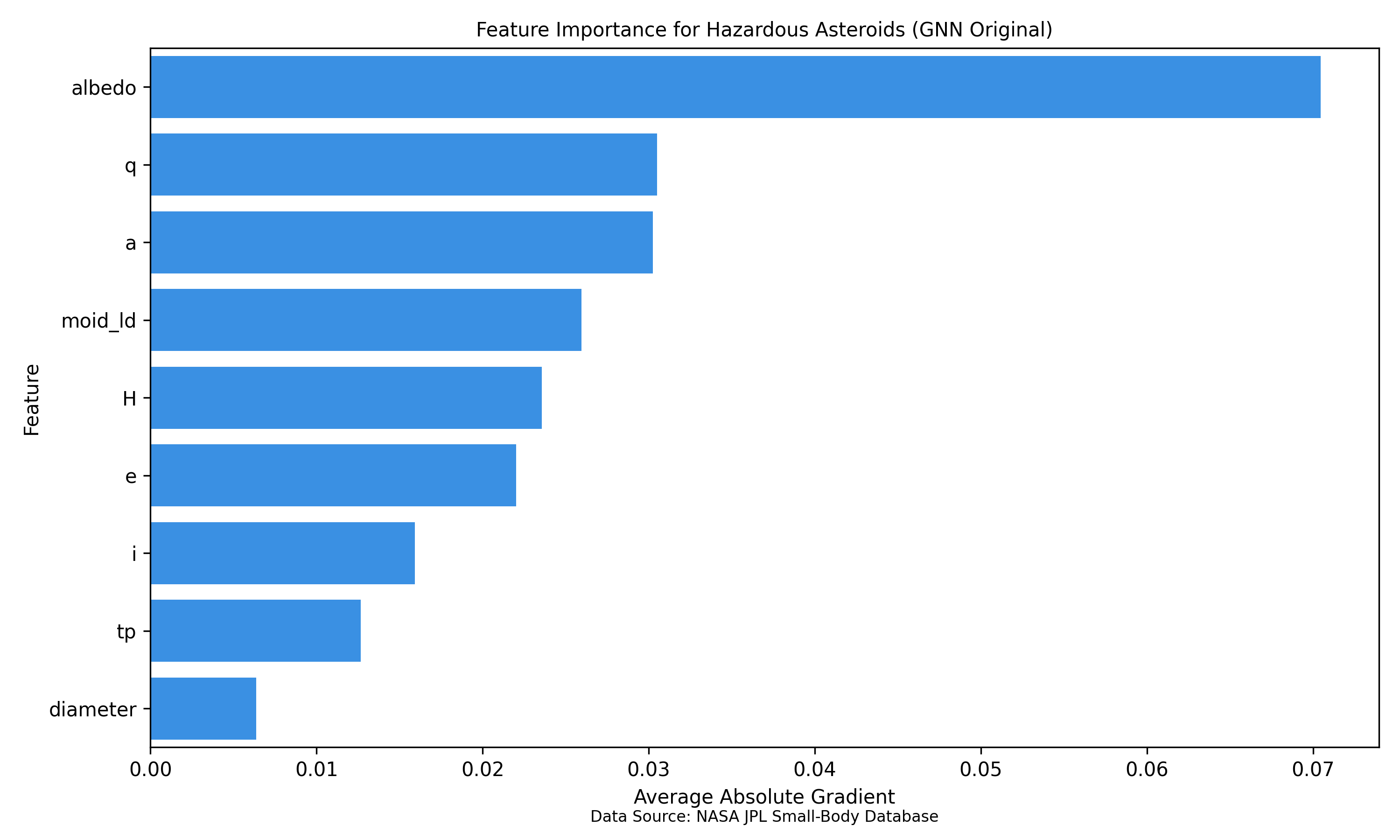}
\caption{Feature Importance for Hazardous Asteroid Classification}
\label{fig:feature_importance}
\end{figure}

Feature importance analysis (Figure \ref{fig:feature_importance}) highlights the most influential features for hazard classification. Albedo emerged as the most significant, likely due to its correlation with an asteroid’s size and composition, which affect its detectability and impact potential. Perihelion distance and semi-major axis also play key roles, which reflects their importance in determining an asteroid’s proximity to Earth’s orbit. Other features like minimum orbit intersection distance, absolute magnitude, and eccentricity contribute to the model’s decision-making, while diameter has the least influence, possibly due to its high missing values rate.

\section{Discussion}

Our work leverages a Graph Neural Network (GNN) to classify potentially hazardous asteroids (PHAs), addressing limitations of traditional machine learning methods that overlook dynamical relationships. Trained on a NASA dataset of 958,524 asteroids, our model navigates extreme class imbalance (0.22\% hazardous) and significant data sparsity, with features like diameter and albedo missing over 85\% of values. This section evaluates our model’s performance, critiques past research’s reliance on accuracy, and discusses implications for planetary defense and deep space navigation, proposing future directions to enhance our approach.

\subsection{Model Performance}

Our GNN model achieved an overall accuracy of 99\% and an Area Under the ROC Curve (AUC) of 0.99 which indicates strong discriminative power. For the hazardous class, with only 0.22\% of the dataset, it attained a precision of 24\%, recall of 78\%, and F1-score of 37\% after applying Synthetic Minority Oversampling Technique (SMOTE), as shown in Table \ref{tab:classification_metrics}. The high recall ensures most hazardous asteroids are identified, critical for planetary defense, though the low precision reflects a high false positive rate. Feature importance analysis in \ref{fig:feature_importance}) highlights albedo (0.0704), perihelion distance (0.0305), and semi-major axis (0.0303) as key predictors, which aligns with their physical relevance to hazard potential.

\begin{table}[H]
\centering
\caption{Classification Performance Metrics for Final GNN Model}
\begin{tabular}{lcccc}
\hline
\textbf{Class} & \textbf{Precision (\%)} & \textbf{Recall (\%)} & \textbf{F1-score} & \textbf{Support} \\
\hline
Non-hazardous (0) & 100 & 99 & 1.00 & 187,308 \\
Hazardous (1) & 24 & 78 & 0.37 & 413 \\
\hline
Accuracy & \multicolumn{3}{c}{99} & 187,721 \\
Macro Avg & 62 & 89 & 0.68 & 187,721 \\
Weighted Avg & 100 & 99 & 1.00 & 187,721 \\
\hline
\end{tabular}
\label{tab:classification_metrics}
\end{table}

\subsection{Class Imbalance}

The dataset’s extreme imbalance (0.22\% hazardous) biases models toward the non-hazardous class, risking poor minority class performance. SMOTE balanced the training set as shown in \ref{fig:after-smote}. This improved the hazardous class F1-score to 37\%, while Focal Loss emphasized hard-to-classify examples to boost the recall to 78\%. Unlike traditional methods that may ignore the minority class, our approach ensures most hazardous asteroids are detected, though precision remains a challenge.

\subsection{Comparison with Past Research}

Prior studies often report high accuracies without addressing class imbalance, inflating performance perceptions. Table \ref{tab:performance_comparison} compares our model with three relevant works. Si (2018) achieved 99.99\% accuracy using Random Forest but lacked specific hazardous class metrics, likely reflecting overall performance on a balanced dataset \cite{si_2018}. Khajuria et al. (2023) reported 91.9\% accuracy with a Random Forest model, but without precision or recall for hazardous asteroids, its minority class performance is unclear. Malakouti (2023)\cite{malakouti_2023} attained an ROC AUC of 94\% for hazardous asteroids using Random Forest, lower than our 99\% AUC. This indicates our GNN’s superior discriminative power. Unlike these studies, our focus on precision, recall, and F1-score for the hazardous class offers a more robust evaluation, critical for real-world applications where missing a PHA is costly.


\subsection{Methodology and Explainability}

Our methodology leverages GNNs to capture complex interdependencies among asteroids, a significant departure from traditional approaches. The graph construction process to define edges based on orbital similarity, enables the model to account for gravitational and kinematic relationships. The GNN architecture, with three convolutional layers, an attention layer, and a classification head, processes these relationships to produce interpretable predictions. Feature importance analysis, as shown in \ref{fig:feature_importance} identifies albedo (0.0704) as the most influential feature, likely due to its correlation with asteroid size and composition, which affect detectability and impact potential. Orbital parameters like perihelion distance (q, 0.0305), semi-major axis (a, 0.0303), and minimum orbit intersection distance also contribute significantly, aligning with their role in determining Earth proximity. This explainability enhances trust in the model’s predictions, a critical factor for applications in space sciences.

\subsection{Implications}
Accurate classification of potentially hazardous asteroids (PHAs) is critical for planetary defense and autonomous deep space navigation. Our results show that artificial intelligence can support proactive planetary defense strategies. By integrating AI-driven asteroid classification into spacecraft navigation systems, these models can enables real-time hazard detection and avoidance, forming a foundational framework for autonomous deep space navigation. For instance, NASA’s NEO Surveyor mission \cite{nasa_neosurveyor}, scheduled for launch in 2026, could leverage AI technologies to enhance the detection and characterization of near-Earth objects. Similarly, ESA’s Ramses mission \cite{esa_ramses}, set to observe asteroid Apophis in 2029, will benefit from autonomous systems capable of navigating asteroid-rich regions. Furthermore, commercial ventures planning asteroid mining operations can utilize our model’s explainable predictions, driven by feature importance to identify valuable targets while ensuring safe trajectories. This work thus provides a robust platform for intelligent, autonomous systems essential for the next generation of space exploration and planetary defense.

\subsection{Future Directions}

To further enhance our model, we propose several directions. First, advanced imbalance handling techniques, such as ensemble methods or adaptive loss functions, could improve precision without sacrificing recall. Second, enhancing explainability through subgraph analysis could identify specific relationships contributing to false positives and refine the model’s decision-making. Finally, hyperparameter optimization and architecture refinements, such as experimenting with Graph Attention Networks (GAT) or GraphSAGE, could further boost performance.


\section{Conclusion}
This research introduced a novel Graph Neural Network (GNN) approach for classifying potentially hazardous asteroids (PHAs), addressing the limitations of traditional machine learning methods that overlook dynamical relationships among asteroids. Our model achieved an overall accuracy of 99\% and an Area Under the ROC Curve (AUC) of 0.99. For the hazardous class, comprising only 0.22\% of the dataset, the model attained a precision of 24\%, recall of 78\%, and F1-score of 37\%. The overall model's F1-score was 0.37. The high recall ensures most PHAs are identified, critical for planetary defense, while feature importance analysis highlighted albedo, perihelion distance, and semi-major axis (0.0303) as key predictors.

Our work advances asteroid classification by leveraging GNNs to model complex interdependencies, offering a more comprehensive and interpretable framework than traditional methods. This approach supports planetary defense by enabling early detection and tracking of PHAs, as demonstrated by its relevance to missions like NASA’s Double Asteroid Redirection Test \cite{nasa_dart} and ESA’s Hera mission \cite{esa_hera}, which focus on asteroid deflection strategies. This framework can be the foundation for autonomous deep space navigation, essential for safe trajectory planning in asteroid-rich regions, as seen in upcoming missions such as NASA’s NEO Surveyor \cite{nasa_neosurveyor} and ESA’s Ramses \cite{esa_ramses}. Future work will aim to enhance precision through advanced imbalance handling techniques and explore sophisticated GNN architectures like GraphSAGE to improve relationship modeling. These advancements will strengthen the framework’s utility for real-world applications and contribute to safer and more efficient planetary defense and space exploration initiatives.

\subsection{Data Availability and Reproduction}

The dataset used in this research, comprising over 950,000 asteroid records from NASA's Jet Propulsion Laboratory (JPL) Small-Body Database, is openly available under the OpenData Commons Open Database License (ODbL) v1.0. The results, experiments, data, and reproduction guide are publicly accessible on GitHub at \url{https://github.com/baimamboukar/hazardous-asteroid-classification} under the MIT license. For detailed instructions on reproducing the experiments, including data preprocessing, model training, and evaluation, please refer to the repository's documentation.

\bibliographystyle{IEEEtran}
\bibliography{citations}
\end{document}